\documentclass[aps,prl,floatfix,showpacs,twocolumn]{revtex4-1}
\usepackage{ulem}
\usepackage{dcolumn}
\usepackage{bm}
\usepackage{color}
\usepackage{amssymb}
\usepackage{amsmath}
\usepackage{graphicx}
\usepackage{amsfonts}
\usepackage{slashed}
\usepackage{pstricks}
\usepackage{float}
\usepackage{hyperref}
\usepackage{array}
\allowdisplaybreaks
\usepackage{dcolumn}
\usepackage{epsf}
\begin{document}

\begin{flushright}
LA-UR-17-28128
\end{flushright}
\title{Quantum Monte Carlo calculations of weak transitions in $A\,$=$\,$6--10 nuclei}
\author{S.\ Pastore$^{\rm a}$, A.\ Baroni$^{\rm b}$, J.\ Carlson$^{\rm a}$, 
S.\ Gandolfi$^{\rm a}$, Steven C.\ Pieper$^{\rm d}$, R.\ Schiavilla$^{\rm b,c}$, and R.B.\ Wiringa$^{\rm d}$}
\affiliation{
$^{\rm a}$\mbox{Theoretical Division, Los Alamos National Laboratory, Los Alamos, NM 87545 }
$^{\rm b}$\mbox{Department of Physics, Old Dominion University, Norfolk, VA 23529}
$^{\rm c}$\mbox{Theory Center, Jefferson Lab, Newport News, VA 23606}\\
$^{\rm d}$\mbox{Physics Division, Argonne National Laboratory, Argonne, IL 60439}
}
\date{\today}
\begin{abstract}
{\it Ab initio} calculations of the  Gamow-Teller (GT) matrix elements in the $\beta$ decays of $^6$He and $^{10}$C
and electron captures in $^7$Be are carried out using both variational and Green's function Monte Carlo
wave functions obtained from the Argonne $v_{18}$ two-nucleon and Illinois-7 three-nucleon interactions, and
axial many-body currents derived from either meson-exchange phenomenology or chiral effective field theory.
The agreement with experimental data is excellent for the electron captures in $^7$Be, while theory overestimates
the $^6$He and $^{10}$C data by $\sim 2\%$  and $\sim 10\%$, respectively.  We show that for these systems
correlations in the nuclear wave functions are crucial to explain the data, while many-body currents increase by
$\sim 2$--$3\%$ the one-body GT contributions.  These findings suggest that the longstanding $g_A$-problem,
{\it i.e.}, the systematic overprediction ($\sim 20 \%$ in $A\le 18$ nuclei) of GT matrix elements in shell-model
calculations, may be resolved, at least partially, by correlation effects.  
\end{abstract}

\maketitle
A major objective of nuclear theory is to explain the structure and dynamics of nuclei
in a fully microscopic approach.  In such an approach the nucleons interact with each 
other in terms of many-body (primarily, two- and three-body)
effective interactions, and with external electroweak probes via effective currents describing the
coupling of these probes to individual nucleons and many-body clusters of them.  We will refer
below to this approach as the {\it basic model} of nuclear theory.

For light nuclei (s- and p-shell nuclei up to $^{12}$C), quantum Monte Carlo (QMC) and, in
particular, Green's Function Monte Carlo (GFMC) methods allow us to carry out first-principles,
accurate calculations of a variety of nuclear properties~\cite{Bacca:2014,Carlson:1998,Carlson:2015} within the basic
model.  These calculations retain the full complexity of the many-body correlations induced by the
Hamiltonians and currents, which have an intricate spin-isospin operator structure.  When
coupled to these numerically accurate QMC methods, the deceptively simple picture put forward
in the basic model provides a quantitative and accurate description of the structure and dynamics
of light nuclei over a broad energy range, from the keV's relevant in nuclear astrophysical
contexts~\cite{Carlson:2015,Marcucci:2006,Adelberger:2011}, to the MeV's of low-lying nuclear
spectra~\cite{Carlson:2015,Piarulli:2017} and radiative decay processes~\cite{Bacca:2014,Pastore:2013}, to the GeV's
probing the short-range structure of nuclei and the limits of the basic model
itself~\cite{Bacca:2014,Schiavilla:2007,Wiringa:2014,Marcucci:2016}.

In the present study we focus on low-energy weak transitions in nuclei with mass number
$A\,$=$\, 6$--10.  To the best of our knowledge, calculations of $\beta$-decays and electron-capture
processes in this mass range have relied so far, with the exception of
Refs.~\cite{Schiavilla:2002,Pervin:2007} discussed below and of Ref.~\cite{Vaintraub:2009}
reporting on the $^6$He $\beta$-decay, on relatively simple shell-model
or cluster descriptions of the nuclear states involved in the transitions.  The shell
model---itself an approximation of the basic model---has typically failed to reproduce
the measured Gamow-Teller (GT) matrix elements governing these weak transitions,
unless use was made of an effective one-body GT operator, in which the nucleon axial coupling
constant $g_A$ is quenched relative to its free value~\cite{Chou:1993,Engel:2017} (ranging
from $g^{\rm eff}_A \simeq 0.85 \, g_A$ in the light nuclei under consideration here to
$g^{\rm eff}_A \simeq 0.7 \, g_A$ in heavy nuclei).
More phenomenological models
have been based on $\alpha$-nucleon-nucleon (for $A$=6)
or $\alpha$-$^3$H and $\alpha$-$^3$He (for $A$=7) or
$\alpha$-$\alpha$-nucleon-nucleon (for $A$=10) clusterization, and have used Faddeev
techniques with a separable representation of the nucleon-nucleon and $\alpha$-nucleon
interaction~\cite{Parke:1978} or the resonating-group method~\cite{Walliser:1983} or
rather crude potential wells~\cite{Bartis:1963}.  While these studies
provide useful insights into the structure of these light systems, nevertheless their connection to the
basic model is rather tenuous.  In particular, they do not explain whether the required quenching of
$g_A$ in shell-model calculations reflects deficiencies in the corresponding wave functions---possibly
due to the lack of correlations and/or to limitations in model space---or in the model adopted for the
nuclear axial current, in which many-body terms are typically neglected.

The first QMC calculation of the $A\,$=$\,6$--7 weak transitions in the
basic model was carried out with the Variational Monte Carlo (VMC) method in
Ref.~\cite{Schiavilla:2002}.   It used nuclear axial currents including, apart from
the (one-body) GT operator, two-body operators, which arise naturally in a
meson-exchange picture ($\pi$- and $\rho$-exchange, and $\rho\pi$-transition
mechanisms) and when excitations of nucleon resonances (notably the $\Delta$
isobar) are taken into account.  These two-body operators, multiplied by hadronic
form factors so as to regularize their short-range behavior in configuration space,
were then constrained to reproduce the GT matrix element contributing to tritium
$\beta$ decay by adjusting the poorly known $N$-to-$\Delta$ axial coupling
constant (see Ref.~\cite{Shen:2012} for a recent summary).

Yet, the calculations of Ref.~\cite{Schiavilla:2002} were based on {\it approximate}
VMC wave functions to describe the nuclear states involved in the transitions.  This
shortcoming was remedied in the subsequent GFMC study of Ref.~\cite{Pervin:2007},
which, however, only retained the one-body GT operator.  Adding to the GFMC-calculated
one-body matrix elements the VMC estimates of two-body contributions obtained in
Ref.~\cite{Schiavilla:2002} led Pervin {\it et al.}~\cite{Pervin:2007} to speculate that
a full GFMC calculation of these $A\,$=$\,$6--7 weak transitions might be in agreement
with the measured values.

The last three decades have witnessed the emergence of chiral effective field theory
($\chi$EFT)~\cite{Weinberg:1990}.  In $\chi$EFT, the symmetries of quantum chromodynamics
(QCD), in particular its approximate chiral symmetry, are used to systematically constrain
classes of Lagrangians describing, at low energies, the interactions of nucleons and
$\Delta$ isobars with pions as well as the interactions of these hadrons with electroweak
fields~\cite{Park:1993,Park:1996}.  Thus $\chi$EFT provides a direct link between QCD
and its symmetries, on one side, and the strong and electroweak interactions in nuclei,
on the other.  Germane to the subject of the present letter are, in particular, the recent
$\chi$EFT derivations up to one loop of nuclear axial currents reported in Refs.~\cite{Baroni:2016,Krebs:2016}.
Both these studies were based on time-ordered perturbation theory and a power-counting scheme
{\it \`a la} Weinberg, but adopted different prescriptions for isolating non-iterative terms in reducible
contributions.  There are differences---the origin of which is yet unresolved---in the loop corrections
associated with box diagrams in these two independent derivations.

\begin{figure}[bth]
 \includegraphics[width=2in]{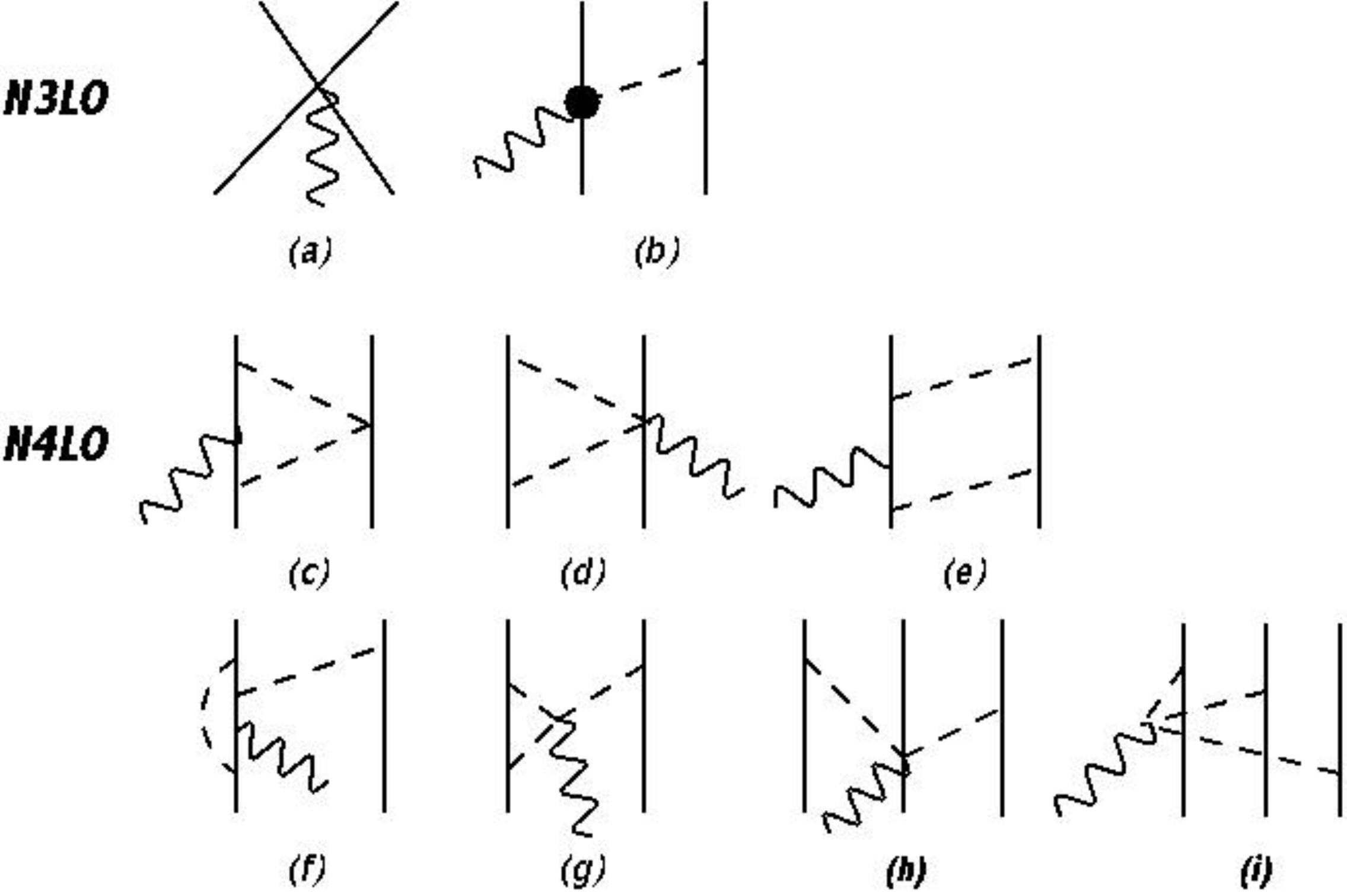}
 \caption{Diagrams illustrating the (non-vanishing) contributions to the 2b and
 3b axial currents.  Nucleons, pions, and external fields are denoted by solid, dashed
 and wavy lines, respectively.  The circle in panel (b) represents the vertex implied by the
 ${\cal L}^{(2)}_{\pi N}$ chiral Lagrangian~\cite{Fettes:2000}, involving the LECs $c_3$ and
 $c_4$.    Only a single time ordering is shown; in particular, all direct- and crossed-box diagrams
 are accounted for.  The power counting of the various contributions is also indicated.
 See text for further explanations.
}
\label{fig:f1}
\end{figure}

The present study reports on VMC and GFMC calculations of weak transitions in $^6$He, $^7$Be,
and $^{10}$C, based on the Argonne $v_{18}$ (AV18) two-nucleon~\cite{Wiringa:1995} and
Illinois-7 (IL7) three-nucleon~\cite{Pieper:2008} interactions, and axial currents obtained either in the
meson-exchange~\cite{Shen:2012} or $\chi$EFT~\cite{Baroni:2016} frameworks mentioned earlier.
The AV18+IL7 Hamiltonian reproduces well the observed spectra of light nuclei ($A\,$=3--12),
including the $^{12}$C ground- and Hoyle-state energies~\cite{Carlson:2015}.
The meson-exchange model for the nuclear axial current has been most recently
reviewed in Ref.~\cite{Shen:2012}, where explicit expressions for the various one-body (1b) and
two-body (2b) operators are also listed (including fitted values of the $N$-to-$\Delta$ axial
coupling constant).  The $\chi$EFT axial current~\cite{Baroni:2016,Baroni:2016a}
consists of 1b,  2b, and three-body (3b) operators.  The 1b
operators read
\begin{equation}
{\bf j}^{\rm 1b}_{5,\pm}=-g_A \sum_{i=1}^A  \tau_{i,\pm} \left( 
{\bm\sigma}_i -\!\frac{ {\bm \nabla}_i\,\,{\bm \sigma}_i\cdot{\bm \nabla}_i 
 - {\bm \sigma}_i \,  \nabla^2_i}{2\, m^2}\right)  ,
\label{eq:gto}
\end{equation}
where $\tau_{i,\pm}=(\tau_{i,x}\pm i\, \tau_{i,y})/2$ is the standard isospin raising ($+$)
or lowering ($-$) operator, and ${\bm \sigma}_i$ and $-i \,{\bm \nabla}_i$ are, respectively,
the Pauli spin matrix and momentum operator of nucleon $i$.  
The 2b and 3b operators are illustrated diagrammatically in Fig.~\ref{fig:f1}
in the limit of vanishing momentum transfer considered here.
Referring to Fig.~\ref{fig:f1}, the 2b operators are from contact [CT, panel (a)],
one-pion exchange (OPE) [panels (b) and (f)], and multi-pion exchange (MPE)
[panels (c)-(e) and (g)],
\begin{equation}
{\bf j}^{\rm 2b}_{5,\pm}=\sum_{i<j=1}^A \Big[ \,   {\bf j}^{\rm CT}_{5,\pm}(ij)
+{\bf j}^{\rm OPE}_{5,\pm}(ij) +{\bf j}^{\rm MPE}_{5,\pm}(ij)\Big] \ ,
\label{eq:2b}
\end{equation}
and the 3b operators are from MPE [panels (h)-(i)],
\begin{equation}
{\bf j}^{\rm 3b}_{5,\pm}=\sum_{i<j<k=1}^A 
{\bf j}^{\rm MPE}_{5,\pm}(ijk) \ .
\label{eq:3b}
\end{equation}
Configuration-space
expressions for these 2b and 3b operators are reported in Ref.~\cite{Baroni:2016a}.

The 1b operator in Eq.~(\ref{eq:gto}) includes the leading order (LO) GT term and the first
non-vanishing corrections to it, which come in at next-to-next-to-leading order (N2LO)~\cite{Baroni:2016a}.
Long-range 2b corrections from OPE enter at N3LO, panel (b) in Fig.~\ref{fig:f1}, involving the
low-energy constants (LECs) $c_3$ and $c_4$ in the sub-leading ${\cal L}^{(2)}_{\pi N}$ chiral
Lagrangian~\cite{Fettes:2000}, as well as at N4LO, panel (f).  
In terms of the expansion parameter $Q/\Lambda_\chi$---where $Q$ specifies generically 
the low-momentum scale and $\Lambda_\chi \!=\!1$ GeV is the chiral-symmetry-breaking 
scale---they scale as $(Q/\Lambda_\chi)^3$
and $(Q/\Lambda_\chi)^4$, respectively, relative to the LO.  Loop corrections from MPE,
panels (c)-(e) and (g), come in at N4LO, as do 3b currents, panels (h)-(i).
Finally, the contact 2b current at N3LO, panel (a), is proportional to a LEC, denoted as $z_0$.

The short-range behavior of the 2b and 3b operators is regularized by
including a cutoff $C_\Lambda(k)\,$=$\,{\rm exp}(-k^4/\Lambda^4)$ in momentum space~\cite{Baroni:2016a},
and the values $\Lambda\!=\!500$ and 600 MeV are considered in the present work.
In correspondence to each $\Lambda$ and to each set of $(c_3,c_4)$, either
$(c_3,c_4)=(-3.2,5.4)$ GeV$^{-1}$ as reported in Ref.~\cite{Machleidt:2011}
or $(c_3,c_4)=(-5.61,4.26)$ GeV$^{-1}$ as determined in Ref.~\cite{Hoferichter:2015},
the LEC $z_0$ is constrained to reproduce the measured GT matrix element of
tritium in hyperspherical-harmonics calculations based on the AV18+UIX~\cite{Pudliner:1995}
Hamiltonian~\cite{Baroni:2016a}. With the AV18+IL7 Hamiltonian adopted here,
the calculated 
tritium GT matrix element is within $\lesssim 1.5\%$ of the experimental datum.


Reduced matrix elements (RMEs) for the $\beta$ decays between the $^6$He($0^+;1$) and
$^6$Li($1^+;0$) ground states, and between the $^{10}$C($0^+;1$) ground state and
$^{10}$B($1^+;0$) first excited state, and $\epsilon$ captures of the $^7$Be($3/2^-;1/2$)
ground state to the $^7$Li($3/2^-;1/2$) ground state
and  $^7$Li($1/2^-;1/2$) first excited state are listed in Table~\ref{tb:tb1}
(in parentheses are the spin-parity, $J^\pi$, and isospin, $T$, assignments for each state).
All processes are allowed or superallowed, and are therefore driven (almost) exclusively by the axial
current (and, additionally, the vector charge---the Fermi operator---for the
transition between the ground states of $^7$Be and $^7$Li).  Retardation effects from
the momentum transfer dependence of the operators, and corrections from suppressed
transitions, such as, for example, those induced in the $A\,$=$\,6$ and 10 decays
by the magnetic dipole associated with the vector current, are negligible~\cite{Schiavilla:2002}.
Therefore the RMEs listed in Table~\ref{tb:tb1} follow simply from
\begin{equation}
{\rm RME}=\frac{\sqrt{2\, J_f+1}}{g_A} \,
\frac{\langle J_f M | j^z_{5,\pm} | J_i M\rangle}{\langle J_iM, 10 |J_f M\rangle} \ ,
\end{equation}
where $j^z_{5,\pm}$ is the $z$-component of the axial current ${\bf j}_{5,\pm}$
(at vanishing momentum transfer) given above and $\langle J_iM, 10 |J_f M\rangle$
are Clebsch-Gordan coefficients.  The VMC results are obtained by straightforward
Monte Carlo integration of the nuclear matrix elements above between (approximate)
VMC wave functions; the GFMC results are from mixed-estimate
evaluations of these matrix elements using previously generated GFMC
configurations for the states under consideration, as illustrated in Ref.~\cite{Pervin:2007}. 
\begin{widetext}
\begin{center}
\begin{table}[bth]
\begin{tabular}{l||c||c|c||c}
                         & $^6$He $\beta$-decay           &$^7$Be $\epsilon$-capture (gs)  & $^7$Be $\epsilon$-capture (ex) & $^{10}$C $\beta$-decay\\
\hline 
${\rm LO}$            &  2.168(2.174)                  &  2.294(2.334)                  & 2.083(2.150)                   & 2.032(2.062) \\
\hline
${\rm N4LO}$           &  3.73(3.03)$\times10^{-2}$      &   6.07(4.98)$\times10^{-2}$    & 4.63(4.63)$\times10^{-2}$      & 1.61(1.55)$\times10^{-2}$ \\
${\rm N4LO}^\star$     &  3.62(3.43)$\times10^{-2}$      &   6.62(5.43)$\times10^{-2}$    & 5.31(5.38)$\times10^{-2}$      & 1.80(1.00)$\times10^{-2}$ \\
\hline
MEC                   &  6.90(4.57)$\times10^{-2}$      &  10.5(10.3)$\times10^{-2}$     & 8.88(8.99)$\times10^{-2}$      &  5.31(4.28)$\times10^{-2}$            \\
\hline
\hline
EXP                      &  2.1609(40)      &  2.3556(47)        & 2.1116(57)     &    1.8331(34)              \\
\hline

\end{tabular}
\caption{
Gamow-Teller RMEs in $A\,$=$\,6$, 7, and 10 nuclei obtained with chiral
axial currents and GFMC (VMC) wave functions corresponding to the AV18+IL7 Hamiltonian model.
Results corresponding to the one-body LO contribution (row labeled LO) and to the sum of all
corrections beyond
LO obtained with cutoff $\Lambda$=500 MeV and 600 MeV (rows labeled respectively as N4LO and
N4LO$^\star$), are listed.  
The sum of all two-body corrections obtained with conventional
meson-exchange axial currents is listed in the row labeled MEC. 
Cumulative contributions, to be compared with the experimental
data~\cite{Knecht:2012,Chou:1993,Bambynek:1977,Towner:2017} reported in the last row,
are obtained by adding to the  LO terms  
the contributions from either the chiral (${\rm N4LO}$ or ${\rm N4LO}^\star$) or
the conventional (MEC) currents. 
Statistical errors associated with the Monte
Carlo integrations are not shown, but are $\sim 1\%$.
}
\label{tb:tb1}
\end{table}
\end{center}
\end{widetext}

The sum of all contributions beyond LO, denoted as N4LO and N4LO$^\star$ in Table~\ref{tb:tb1},
leads approximately to a 2--3\% increase in the LO prediction 
for the GT matrix elements of all processes under consideration.  There is some
cutoff dependence in these contributions, as indicated by the difference between the
rows labeled N4LO and N4LO$^\star$ in Table~\ref{tb:tb1}, which may be
aggravated here by the lack of consistency between the $\chi$EFT currents and the
phenomenological potentials used to generate the wave functions, {\it i.e.}, by
the mismatch in the short-range behavior of potentials and currents.  The N4LO and
N4LO$^\star$ results in Table~\ref{tb:tb1} correspond to the set
$(c_3,c_4)\,$=(--3.2,5.4) GeV$^{-1}$~\cite{Machleidt:2011}
in the OPE GT operator at N3LO.  To illustrate the sensitivity of predictions to the set of
$(c_3,c_4)$ values, we observe that use of the more recent determination
$(c_3,c_4)\,$=(--5.61,4.26) GeV$^{-1}$~\cite{Hoferichter:2015} would lead
to an N4LO GFMC-calculated value of $6.71(2.89) \times 10^{-2}$ for the $^7$Be $\epsilon$ capture
to the $^7$Li ground (first excited) state for the choice of cutoff $\Lambda=500$ MeV,
to be compared to the corresponding $6.07(4.63)\times10^{-2}$ reported
in Table~\ref{tb:tb1}.  Lastly, the N4LO contributions obtained with the more accurate
GFMC wave functions are about 20\% larger than those corresponding to
VMC wave functions for the $^6$He and $^7$Be-to-$^7$Li
ground-state transitions, albeit it should be emphasized that this
is in relation a small overall $\sim$ 2\% correction from 2b and 3b operators.

\begin{center}
\begin{table}[bth]
\begin{tabular}{c||c|c}
  &     gs  & ex \\
\hline 
LO                 &   2.334    &  2.150 \\
\hline
N2LO            &    --3.18$\times10^{-2}$    & --2.79$\times10^{-2}$\\
\hline
N3LO(CT)    &     2.79$\times10^{-1}$    &  2.36$\times10^{-1}$     \\
OPE  &    --2.99$\times10^{-2}$   & --2.44$\times10^{-2}$ \\
\hline
N4LO(2b)      &    --1.61$\times10^{-1}$    & --1.33$\times10^{-1}$     \\
N4LO(3b)      &     --6.59$\times10^{-3}$   &   --4.86$\times10^{-3}$    \\
\hline
\end{tabular}
\caption{
Individual contributions to the $^7$Be $\epsilon$-capture Gamow-Teller RMEs obtained at
various orders in the chiral expansion of the axial current ($\Lambda\,$=$\,$500 MeV) with VMC
wave functions.  The rows labeled LO and N2LO refer to, respectively, the first term and the
terms proportional to $1/m^2$ in Eq.~(\ref{eq:gto}); the rows labeled N3LO(CT) and OPE, and
N4LO(2b) and N4LO(3b), refer to panel (a) and panels (b) and (f), and to panels (c)-(e), (g) and
panel (h) in Fig.~\ref{fig:f1}, respectively.
}
\label{tb:tb2}
\end{table}
\end{center}
The contributions of the axial current order-by-order in the chiral expansion
are given for the GT matrix element of the $^7$Be $\epsilon$ capture in
Table~\ref{tb:tb2}.  Those beyond LO, with the exception of
the CT at N3LO, have opposite sign relative to the (dominant) LO.  The
loop corrections N4LO(2b) are more than a factor 5 larger (in magnitude)
than the OPE.  This is primarily due to the accidental cancellation between the
terms proportional to $c_3$ and $c_4$ in the OPE operator at N3LO (which
also occurs in the tritium GT matrix element~\cite{Baroni:2016a}).  It is also in line
with the {\it chiral filter hypothesis}~\cite{Kubodera:1978,Rho:1991,Rho:2008},
according to which, if soft-pion processes are suppressed---as is the case for
the axial current---then higher-order chiral corrections are not necessarily
small.   Indeed, the less than 3\% overall correction due to terms beyond LO reported in
Table~\ref{tb:tb1} (row N4LO) comes about because of destructive interference
between two relatively large ($\sim 10\%$) contributions from the CT and the
remaining [primarily N4LO(2b)] terms considered here. 

 \begin{figure}[bth]
 \includegraphics[width=3.5in]{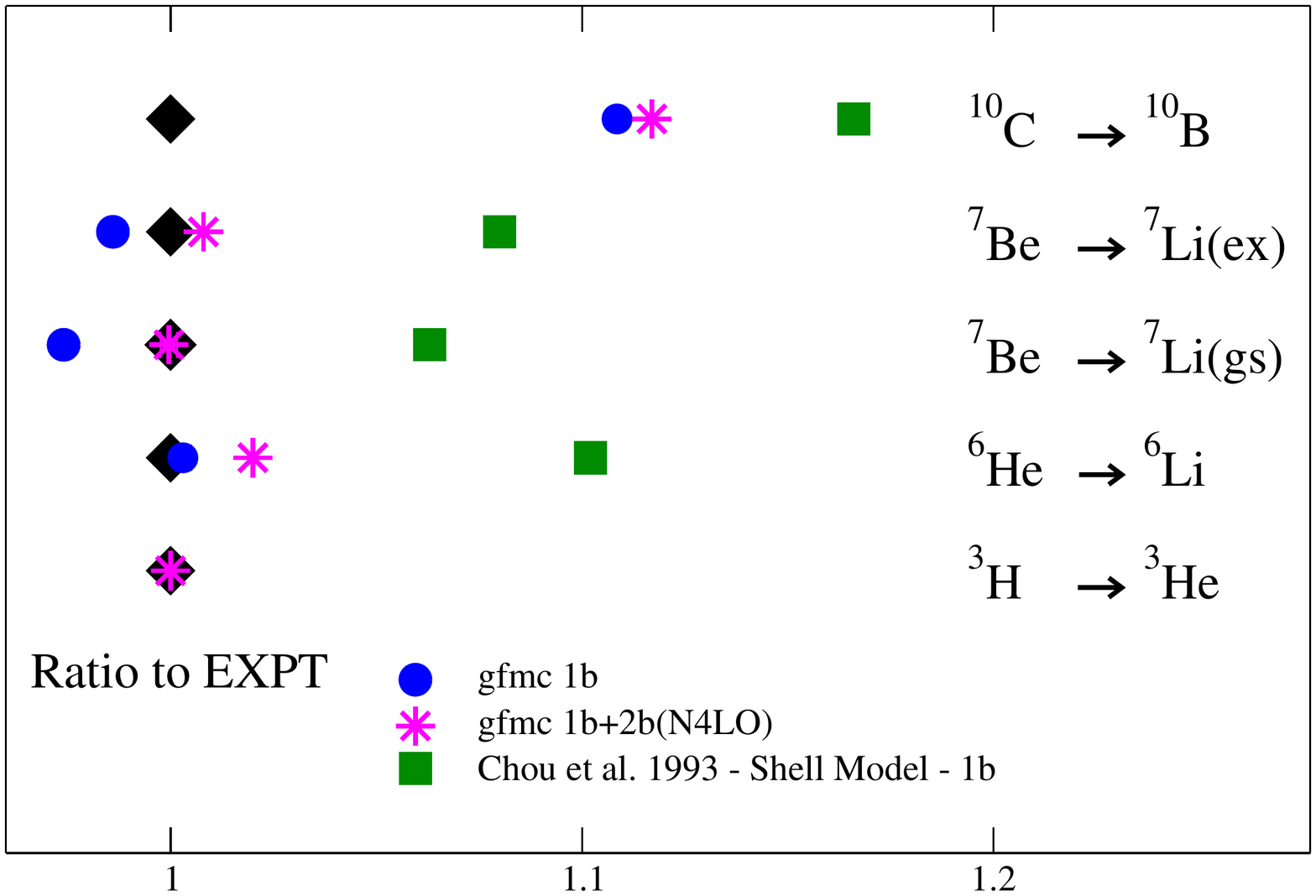}
 \caption{
 (Color online) Ratios of GFMC to experimental values
 of the GT RMEs in the $^3$H, $^6$He, $^7$Be, and $^{10}$C weak transitions.  Theory
 predictions correspond to the $\chi$EFT axial current in LO (blue circles) and up to 
 N4LO (magenta stars). Green squares indicate `unquenched' shell model calculations 
 from Ref.~\cite{Chou:1993} based on the LO axial current. 
 }
\label{fig:f2}
\end{figure}
Ratios of GFMC to experimental values for the GT RMEs in the $^3$H, $^6$He, $^7$Be,
and $^{10}$C weak transitions are displayed in Fig.~\ref{fig:f2}---theory results correspond
to $\chi$EFT axial currents at LO and including corrections up to N4LO.  The experimental
values are those listed in Table~\ref{tb:tb1}, while that for $^3$H is 1.6474(24)~\cite{Baroni:2016a}.
These values have been obtained by using $g_A\,$=$\,1.2723(23)$~\cite{PDG:2016} and
$K/\left[G^2_V\left(1+\Delta^V_R\right)\right]\,$=$\,6144.5(1.4)$ sec~\cite{Hardy:2015},
where $K\,$=$\,2\,\pi^3 \,{\rm ln} \,2/m^5_e\,$=$\, 8120.2776(9) \times 10^{-10}$ GeV$^{-4}\,$sec
and $\Delta^V_R=2.361(38)\%$ is the transition-independent radiative correction~\cite{Hardy:2015}.
In the case of the $\beta$ decays, but not for the $\epsilon$ captures, the transition-dependent
($\delta^\prime_R$) radiative correction has also been accounted for.  Lastly, in
the $\epsilon$ processes the rates have been obtained by ignoring the factors $B_K$ and $B_{L1}$
which include the effects of electron exchange and overlap in the capture from the $K$ and $L1$
atomic subshells.  As noted by Chou {\it et al.}~\cite{Chou:1993} following
Bahcall~\cite{Bahcall:1963,Bahcall:1978}, such an approximation is expected to be
valid in light nuclei, since these factors only account for a redistribution of the total 
strength among the different subshells (however, it should be noted that $B_K$ and $B_{L1}$
were retained in Ref.~\cite{Schiavilla:2002}, and led to the extraction of experimental values
for the GT RMEs about 10\% larger than reported here).

We find overall good agreement with data for the $^6$He $\beta$-decay and $\epsilon$ captures
in $^7$Be, although the former is overpredicted by $\sim 2\%$, a contribution that comes almost 
entirely from 2b and 3b chiral currents. 
The experimental GT RME for the $^{10}$C $\beta$-decay is overpredicted by $\sim 10\%$,
with two-body currents giving a contribution that is comparable to the statistical 
GFMC error. The presence of a second ($1^+;0$) excited state at $\sim 2.15$ MeV 
can potentially contaminate the wave function of the $^{10}$B excited
state at $\sim 0.72$ MeV, making this the hardest transition to calculate reliably. 
 In fact, a small admixture of the second excited state ($\simeq 6\%$ in probability) 
 in the VMC wave function brings the VMC reduced matrix element in statistical
 agreement with the the measured value, a variation that does not
 spoil the overall good agreement we find for the reported
 branching ratios of 98.54(14)$\%$ ($<0.08\%$) to the first (second) $(1^+,0)$
 state of $^{10}$B~\cite{Chou:1993}. 
 Because of the small energy difference of these two levels,
 it would require an expensive GFMC calculation to see
 if this improvement remains or is removed; in lighter
 systems we have found that such changes of
 the trial VMC wave function are removed by GFMC.

%

%

We note that correlations in the wave functions significantly reduce the 
matrix elements, a fact that can be appreciated by comparing the LO GFMC (blue
circles in Fig.~\ref{fig:f2}) and the LO shell model calculations (green squares in
the same figure) from Ref.~\cite{Chou:1993}. Moreover, preliminary variational Monte 
Carlo studies, based on the Norfolk two- and 
three-nucleon chiral potentials~\cite{Piarulli:2017,Piarulli:2015,Piarulli:2016} and the 
LO GT operator, bring the  $^{10}$C prediction only $\sim 4\%$ above 
the experimental datum~\cite{Piarulli.private}, indicating that the $\sim 10\%$
discrepancy we find here may indeed be attributable to deficiencies in the AV18+IL7 wave functions
of $A=10$ nuclei.  

%
In the present study we have shown that weak transitions in $A\,$=$\,$6--10 
nuclei can be satisfactorily explained in the basic model, without having to ``quench'' $g_A$.
Clearly, in order to resolve the mismatch in the short-range behavior between potentials and currents
alluded to earlier, GFMC calculations based on the Norfolk chiral potentials of Refs.~\cite{Piarulli:2017,Piarulli:2016}
and consistent chiral currents are in order.  Work along these lines is in progress.
%

%

Correspondence with I.S.\ Towner in reference to radiative corrections in the $A\,$=$\,$6--10
weak transitions is gratefully acknowledged. S.P. wishes to thank A.\ Hayes 
for her guidance and numerous consultations on branching ratios in $A\,$=10 decays.
The work of S.P., J.C., S.G., S.C.P., and~R.B.W.~has been supported by the NUclear
Computational Low-Energy Initiative (NUCLEI) SciDAC project. This research
is also supported by the U.S.~Department of Energy, Office of Science,  Office of
Nuclear Physics, under contracts DE-AC05-06OR23177 (R.S.),
DE-AC02-06CH11357 (S.C.P.~and R.B.W.), and DE-AC52-06NA25396 and 
Los Alamos LDRD program (J.C.~and S.G.).
Computational resources have been provided by Los Alamos Open
Supercomputing, and
Argonne's Laboratory Computing Resource Center.
We also used resources provided by NERSC, which is supported by the US
DOE under Contract DE-AC02-05CH11231.

\end{document}